\documentclass[12pt]{article}
\usepackage{amsfonts,graphicx,amssymb,amsmath,amsthm,epsfig}
\usepackage{float}
\allowdisplaybreaks
\begin{document}
\title{\textbf{Non-Interacting New Agegraphic Dark Energy Model in $f(Q)$ Gravity}}
\author{M. Sharif$^{1,2}$\thanks {msharif.math@pu.edu.pk}~ and
Madiha Ajmal$^1$ \thanks {madihaajmal222@gmail.com} \\
$^1$ Department of Mathematics and Statistics, The University of Lahore,\\
1-KM Defence Road Lahore-54000, Pakistan.\\
$^2$ Research Center of Astrophysics and Cosmology, Khazar University,\\
Baku, AZ1096, 41 Mehseti Street, Azerbaijan.}

\date{}
\maketitle
\begin{abstract}
In this study, we explore the reconstruction of a new agegraphic
dark energy model in a flat Friedmann-Robertson-Walker spacetime by
$f(Q)$ gravity framework, where $Q$ represents non-metricity. We
assume that the scale factor follows a power-law and explore how
this model aligns with the expanding universe. In this perspective,
we develop a new agegraphic $f(Q)$ model and analyze the graphical
behavior for cosmic evolution. We analyze physical characteristics
of the model using the equation of state parameter,
$(\omega_{D}-\omega^{\prime}_{D})$ and the $(r-s)$ planes. The
equation of state parameter indicates a quintessence era
characterized by accelerated expansion. The
$(\omega_{D}-\omega^{\prime}_{D})$-plane identifies the freezing
region and the Chaplygin gas model is represented in the
$(r-s)$-plane. Finally, we examine the stability of the
non-interacting model by evaluating the squared speed of sound. Our
findings show that the non-interacting new agegraphic dark energy
model effectively resolves the cosmic coincidence problem.
\end{abstract}

\section{Introduction}

The general theory of relativity (GR) has explained many features of
the universe through a range of observational evidence. Recent
observations in astronomy \cite{1}, especially of supernovae type-Ia
(SNeIa) \cite{2}, suggest that the cosmos is currently expanding at
an accelerated pace. Other evidences such as large-scale structure
data \cite{3} and cosmic microwave background (CMB) measurements
\cite{4}, also support this idea. The most accepted explanation for
this acceleration is dark energy (DE), an unknown force with strong
negative pressure driving the expansion. Although the exact nature
of DE remains unclear, it presents a significant challenge for
physicists. Einstein's cosmological constant $\Lambda$ is considered
the main candidate for DE. The simplest model, known as $\Lambda$CDM
(Cold Dark Matter), combines this cosmological constant with CDM.
While the $\Lambda$CDM model effectively explains the universe
acceleration and fits current observational data, it faces two key
issues, namely the cosmic coincidence problem and the fine-tuning
problem \cite{5}. To address the limitations of GR, many modified
gravity theories (MGTs) have been proposed. Two main approaches have
emerged: one introduces DE with strong negative pressure within GR
\cite{3}, while the other modifies or extends GR itself. Recent
experiments show that MGTs can explain the early universe events
like inflation and the later stages of accelerated expansion.

General relativity is based on Riemannian geometry, which uses the
Levi-Civita (LC) connection \cite{6} to ensure that spacetime
follows the rule of metric compatibility. In this framework, both
torsion $\mathcal{T}$ and non-metricity are set to zero, making
curvature the main element that describes gravity. However, by
modifying these assumptions, we can explore alternative gravity
theories, allowing non-metricity, curvature and torsion to coexist.
The teleparallel equivalent of GR (TEGR) \cite{7} emerges when using
a connection that permits torsion but excludes curvature and
non-metricity, where torsion describes gravitational interactions.
Similarly, by focusing on a spacetime with non-metricity but no
torsion or curvature, we arrive at the symmetric teleparallel
formulation of GR (STGR) \cite{8}. Both TEGR and STGR \cite{9}
provide alternative ways of describing gravity while still aligning
with GR's results. In this context, we focus on STGR, where gravity
is governed by non-metricity without the presence of curvature or
torsion. In this approach, the action is designed to produce the
STGR, with a non-metricity scalar ($Q$) analogous to the Ricci
scalar ($R$) in GR. By extending this theory to $f(Q)$ \cite{10}, we
generalize the action, resulting in modified Friedmann equations
that offer a new way to explain the universe accelerated expansion
in purely geometric terms \cite{11}.

Researchers are increasingly interested in exploring non-Riemannian
geometry, such as the $f(Q)$ theory \cite{12}. Lazkoz et al.
\cite{14} looked into the observational limits of $f(Q)$ gravity.
Frusciante \cite{15} introduced a specific model within the $f(Q)$
gravity framework, which closely resembles the $\Lambda$CDM model at
a fundamental level but shows unique features, allowing for
different measurable phenomena. Koussour et al. \cite{16} studied
how cosmic parameters behave in $f(Q)$ gravity. Goswami et al.
\cite{17} explored Friedmann-Robertson-Walker (FRW) cosmology in
Weyl-type $f(Q)$ gravity, showing a transition from deceleration to
acceleration, with numerical solutions that aligned well with
observational data. Bhar \cite{18} proposed a spherically symmetric
quintessence DE model in $f(Q)$ gravity, exploring the interplay of
physical parameters and the viability of the stellar model.
Kiroriwal et al. \cite{19} investigated wormhole solutions in
non-linear $f(Q)$ gravity, focusing on energy conditions and the
role of the $f(Q)$ function in maintaining wormholes. Sadatian and
Hosseini \cite{20} studied a modified $f(Q)$ gravity theory with a
scalar field non-minimally coupled to matter, focusing on an
inflationary model. Recent studies \cite{2k} have explored the
geometry and physics of this gravity, offering key insights into its
basics.

Saha and Rudra \cite{a1} studied the reconstruction of $f(Q)$
gravity from a holographic perspective, developing cosmologically
viable solutions using DE models and analyzing their energy
conditions. Goswami and Das \cite{a2} introduced a parametric form
of DE pressure and compared it with the $\Lambda$CDM model using
observational data. Mhamdi et al. \cite{a3} explored the same
gravity with a power-law model, analyzing cosmic dynamics,
observational constraints and growth index evolution, revealing
quintessence or phantom-like behavior. Enkhili et al. \cite{a4}
studied a dynamical DE model in $f(Q)$ gravity, favoring the
quintessence regime and affecting future cosmic expansion. Mhamdi et
al. \cite{a5} used observational analysis to show that the
exponential model statistically outperforms the power-law and
$\Lambda$CDM models. Sahlu and Abebe \cite{a6} studied late-time
acceleration and large-scale structures in the same gravity theory.

Smitha et al. \cite{a7} analyzed anisotropy dependent non singular
solutions and compact object stability. Wang et al. \cite{a8}
analyzed solar syste, constraints and derived Schwarzchild de sitter
like solutions. El Ouardi et al. \cite{a9} used genetic algorithms
to model independent reconstruction of the $f(Q)$ gravity model,
showing minor deviations from $\Lambda$CDM. Rani et al \cite{a10}
studied the effects of charge and evaluated compact star stability
in the same gravity theory. Shuikla er al \cite{a11} explored
modified Chaplygin gas solutions in $f(Q)$ gravity, analyzing their
impact on the cosmic evolution, energy dynamics and the universe
accelerated expansion. Dubey et al. \cite{a12} investigated a
modified same gravity model as the driver of DE evolution, analyzing
its dynamics, statefinder diagnostics and observational consistency,
revealing Chaplygin gas and quintessence-like behaviors. Kumar et
al. \cite{a13} explored anisotropic models for compact stars in
$f(Q)$ gravity, solving Einstein equations and analyzing mass-radius
relationships and stability for observed pulsars. Dimakis et al.
\cite{a14} derived new solutions for static, spherically symmetric
spacetimes in the same gravity and analyzed their physical
properties and general relativistic limit.

The agegraphic DE (ADE) model is grounded in the interplay between
quantum mechanics and GR, focusing on the fluctuations of spacetime.
It begins with the concept that physical measurements are limited by
the uncertainty principle, which affects how we determine quantities
like distance in Minkowski spacetime. This leads to the
identification of energy density, which is influenced by the age of
the universe. The original ADE model \cite{23} incorporates the age
of the universe into its energy density expression, revealing a
relationship that connects cosmic expansion with accelerated growth.
Energy density is given by the equation
$\varrho_{D}=3n^{2}M^{2}_{p}\mathbb{T}^{-2}$, where $\mathbb{T}$
represents the age of the cosmos and $M^{2}_{p}$ is the Planck mass.
However, this model faced difficulties in explaining the universe's
matter-dominated era. To address these issues, Wei and Cai \cite{24}
proposed the new agegraphic dark energy (NADE) model, which replaces
the universe's age with conformal time, represented as
\begin{equation}\label{1}
\varrho_D= \frac{3n^{2}M^{2}_{p}}{\zeta^{2}}.
\end{equation}
To simplify the subsequent calculations, we set $M^{2}_{p}=1$ and
apply the condition $n>1$, resulting in
\begin{equation}\label{2}
\varrho_D= \frac{3n^2}{\zeta^{2}},
\end{equation}
where $\zeta$ is defined as
\begin{equation}\label{3}
\zeta = \int_{0}^{t} \frac{dt}{a},
\end{equation}
and $a$ denotes the scale factor. This adjustment offers a more
effective framework that could potentially resolve issues like the
coincidence problem \cite{25}. It provides a better fit with
observational data while maintaining a similar energy density
structure as the holographic DE (HDE) model, despite their
fundamental differences. The NADE model stands out for its promising
alignment with cosmic observations and significantly contributes to
our understanding of DE role in the universe evolution.

The reconstruction approach or establishing correspondences between
different DE models, has recently gained significant interest in
cosmology. Zhang et al. \cite{26} examined the NADE model
interaction with matter using the statefinder diagnostic,
illustrating its impact on the universe evolution through plotted
trajectories in the statefinder parameter space. Li et al. \cite{27}
showed that the NADE could be described by a tachyon scalar field,
allowing for the reconstruction of its potential and dynamics. Jamil
and Saridakis \cite{28} investigated the NADE scenario in the
Hoava-Lifshitz gravity. The analysis indicated compatibility with
observations, but it did not clarify potential conceptual and
theoretical issues related to this gravity theory. Sheykhi and Jamil
\cite{29} investigated the cosmological applications of interacting
HDE and NADE within the context of Brans-Dicke theory. Karami et al.
\cite{31} analyzed the NADE model and DM in a non-flat cosmos under
Horava-Lifshitz cosmology, deriving equations for the density and
deceleration parameters.

Setare et al. \cite{33} analyzed structure formation in the NADE
model, finding that structures in this model were smaller and denser
compared to the $\Lambda$CDM models. Saba and Sharif \cite{34}
focused on reconstructing the NADE model in the $f(G,T)$ gravity
framework (where $G$ and $T$ represent the Gauss-Bonnet invariant
and the trace of the energy-momentum tensor (EMT)) and analyzed its
cosmic behavior using diagnostic parameters. Pourbagher and Amani
\cite{35} derived energy density from the NADE model and studied the
$f(\mathcal{T},B)$ gravity model (where $B$ represents the boundary
term), analyzing the universe components, fitting the Hubble
parameter to observational data and investigating model stability.
Sobhanbabu and Santhi \cite{36} studied the NADE and DM within the
Brans-Dicke scalar-tensor theory, analyzing cosmological parameters
and stability. Pinki and Kumar \cite{37} examined a cosmological
model with interacting NADE and DM in Brans-Dicke gravity, showing
adherence to the generalized second law of thermodynamics. Kumar et
al. \cite{38} presented a novel DE scenario based on Kaniadakis
entropy, investigating the cosmological characteristics, evolution
and stability of the model.

The novelty of our work lies in presenting a method to naturally
understand cosmic acceleration by modifying spacetime geometry.
Unlike other modified gravity theories, which rely on extra fields
or fine-tuning $f(Q)$ gravity offers a flexible cosmological model.
Its second-order field equations make calculations simpler, making
it more practical for both theoretical and computational studies.
Another significant advantage is the theory compatibility with
observational data. Studies comparing $f(Q)$ models to cosmological
observations including CMB, SNeIa and baryon acoustic oscillations
(BAO) have demonstrated that it provides a viable alternative to
General Relativity. Some models even show better alignment with
late-time cosmic acceleration, strengthening its potential as a
replacement for $\Lambda$CDM. However, our study is the first to
systematically reconstruct the non-interacting NADE model within the
$f(Q)$ gravity framework. Some physicists suggest that the
interaction between DE and DM could help better understand and solve
the coincidence problem. This is why the study of the
non-interacting NADE model is essential.

In this study, we explore cosmic evolution and the current expansion
within the $f(Q)$ gravity framework using the non-interacting NADE
model. The structure of the paper is organized as follows. In
section \textbf{2}, we provide an overview of $f(Q)$ gravity and its
significance in cosmology. Section \textbf{3} focuses on the FRW
universe, emphasizing the non-interacting nature of DM and DE. In
section \textbf{4}, we examine cosmic behavior and assess the model
stability. The last section summarizes our key findings.

\section{Geometrical Foundation}

In GR, a torsion-free connection that works with the metric leads to
the LC connection \cite{39}. However, it is possible to introduce
two 3-rank tensors, the torsion tensor
$(T_{\alpha\beta}^{\lambda})$, related to the asymmetric part of the
connection and the non-metricity tensor $(Q_{\beta\alpha\sigma})$,
which describe how the metric changes during covariant
differentiation as follows
\begin{equation}\label{4}
T_{\alpha\beta}^{\lambda}=2\hat{\Gamma}^{\lambda}_{[\alpha\beta]},~~
Q_{\beta\alpha\sigma}=\nabla_{\sigma}g_{\beta\alpha}\neq 0.
\end{equation}
In such a spacetime, the asymmetric tensor \cite{40} can be
expressed as
\begin{equation}\label{5}
\hat{\Gamma}^{\lambda}_{\alpha\beta}={\Gamma}^{\lambda}_{\alpha\beta}
+\mathbb{C}^{\lambda}_{\;\alpha\beta}+\mathbb{L}^{\lambda}_{\;\alpha\beta},
\end{equation}
here,
$\mathbb{C}^{\lambda}_{\;\alpha\beta}=\hat{\Gamma}^{\lambda}_{[\alpha\beta]}
+g^{\lambda\sigma}g_{\alpha\kappa}\hat{\Gamma}^{\kappa}_{[\beta\sigma]}
+g^{\lambda\sigma}g_{\beta\kappa}\hat{\Gamma}^{\kappa}_{[\alpha\sigma]}$
is the contortion tensor,
$\mathbb{L}^{\lambda}_{\;\alpha\beta}=\frac{1}{2}g^{\lambda\sigma}(Q_{\beta\alpha\sigma}
+Q_{\alpha\beta\sigma}-Q_{\lambda\alpha\beta})$ is the disformation
tensor and the LC connection is denoted by
\begin{equation}\label{6}
\Gamma^{\lambda}_{\alpha\beta}=\frac{1}{2}g^{\lambda\sigma}
(g_{\sigma\alpha,\beta}+g_{\sigma\beta,\alpha}-g_{\alpha\beta,\sigma}).
\end{equation}
The LC connection in symmetric connections can be represented using
the disformation tensor as
\begin{equation}\label{7}
\Gamma^{\lambda}_{\beta\alpha}=-\mathbb{L}^{\lambda}_{\;\beta\alpha}.
\end{equation}
The gravitational action in a non-covariant form is given by
\begin{equation}\label{8}
S=\frac{1}{2}\int g^{\beta\alpha}(\Gamma^{\mu}_{\sigma\beta}
\Gamma^{\sigma}_{\alpha\mu} -\Gamma^{\mu}_{\sigma\mu}
\Gamma^{\sigma}_{\beta\alpha})\sqrt{-g} d^ {4}x.
\end{equation}
Using the relation above, the action integral becomes
\begin{equation}\label{9}
S=-\frac{1}{2} \int g^{\beta\alpha}(\mathbb{L}^{\mu}_{~\sigma\beta}
\mathbb{L}^{\sigma}_{~\alpha\mu} - \mathbb{L}^{\mu}_{~\sigma\mu}
\mathbb{L}^{\sigma}_{~\beta\alpha}) \sqrt{-g} d^ {4}x,
\end{equation}
this is the STGR.

We explore an extension of the STGR Lagrangian as follows
\begin{equation}\label{10}
S=\int\bigg(\frac{1}{2}f(Q)+\mathcal{L}_{\mathbf{m}}\bigg)
\sqrt{-g}d^{4}x,
\end{equation}
where $\mathcal{L}_{\mathbf{m}}$ represents the lagrangian density
of matter and $g$ is the determinant of the metric tensor. The
superpotential is defined as
\begin{equation}\label{11}
\mathbb{P}^{\mu\alpha\beta}=\frac{1}{4}\big[-Q^{\mu\alpha\beta}+Q^{\alpha\mu\beta}
+Q^{\beta\mu\alpha}+Q^{\alpha\mu\beta}-\tilde{Q}_{\mu}g^{\alpha\beta}
+Q^{\mu}g^{\alpha\beta}\big],
\end{equation}
where
\begin{equation}\nonumber
Q_{\mu}=Q^{~\alpha}_{\mu~\alpha},\quad
\tilde{Q}_{\mu}=Q^{\alpha}_{~\mu\alpha}.
\end{equation}
The non-metricity is calculated as \cite{41}
\begin{equation}\label{12}
Q=-Q_{\mu\beta\alpha}\mathbb{P}^{\mu\beta\alpha}=-\frac{1}{4}(-Q^{\mu\alpha\rho}Q_{\mu\alpha\rho}
+2Q^{\mu\alpha\rho}Q_{\rho\mu\alpha}-2Q^{\rho}\tilde{Q}_{\rho}+Q^{\rho}Q_{\rho}).
\end{equation}
The field equations are derived
\begin{equation}\label{13}
\frac{-2}{\sqrt{-g}}\nabla_{\mu}(f_{Q}\sqrt{-g}
P^{\mu}_{~\beta\alpha})-\frac{1}{2}f g_{\beta\alpha}-f_{Q}
(P_{\beta\mu\nu}Q_{\alpha}^{~\mu\nu}-2Q^{\mu\nu}_{~~~\beta}
P_{\mu\nu\alpha})= T_{\beta\alpha},
\end{equation}
where $f_{Q}=\frac{\partial f(Q)}{\partial Q}$.

\section{The FRW Universe Model}

The metric for a spatially homogeneous and isotropic universe model
is expressed as
\begin{equation}\label{14}
ds^{2}=a^{2}(t)(dx^{2}+dy^{2}+dz^{2})-dt^{2}.
\end{equation}
The total EMT for DM and DE is expressed as
\begin{equation}\label{15}
T_{\alpha\beta}=T^{DM}_{\alpha\beta}+T^{DE}_{\alpha\beta},
\end{equation}
where
$T^{DM}_{\alpha\beta}=(\varrho_{\mathbf{m}})u_{\alpha}u_{\beta}$
represents the EMT for pressureless DM and
$T^{DE}_{\alpha\beta}=(\varrho_{D}+\mathbb{P}_{D})u_{\alpha}u_{\beta}+
\mathbb{P}_{D}g_{\alpha\beta}$ is the EMT for DE. Here, $u_{\alpha}$
denotes the four-velocity, $\mathbb{P}_{D}$ is the pressure of DE,
and $\varrho_{\mathbf{m}}$ and $\varrho_{D}$ are the energy
densities of DM and DE, respectively. The energy densities for DE
and DM can be represented in fractional form as
\begin{equation}\label{16}
\Omega_{\mathbf{m}}=\frac{\varrho_{\mathbf{m}}}{\varrho_{cr}}=
\frac{\varrho_{\mathbf{m}}}{3\emph{H}^{2}}, \quad
\Omega_{D}=\frac{\varrho_{D}}{\varrho_{cr}}=\frac{\varrho_{D}}{3\emph{H}^{2}},
\end{equation}
indicating that $1$ can be expressed as the sum of $\Omega_{D}$ and
$\Omega_{\mathbf{m}}$, with $\varrho_{cr}$ being the critical
density. For non-interacting DM and DE, the continuity equations are
given as
\begin{eqnarray}\label{17}
\dot{\varrho}_{\mathbf{m}}+3\emph{H}(\varrho_{\mathbf{m}})&=&0,\\\label{18}
\dot{\varrho}_{D}+3\emph{H}(\varrho_{D}+\mathbb{P}_{D})&=&0.
\end{eqnarray}

Assuming the scale factor follows a power-law form
\begin{equation}\label{19}
a(t)=a_{0}t^{m},
\end{equation}
where $m$ and $a_{0}$ are constants and $a_{0}$ is normalized to
$1$. We choose the power-law form for the scale factor for several
reasons. Mathematically, it simplifies the field equations, making
them more tractable within the modified gravity framework,
particularly in $f(Q)$ gravity and facilitates a clearer
understanding of cosmic evolution. This form allows direct solutions
of the field equations, which are crucial for analyzing the universe
expansion. Power-law cosmology is widely used in gravitational
models as it effectively represents different cosmic phases
including radiation-dominated, matter-dominated and accelerated
expansion eras. Its smooth transition between these phases provides
a robust framework for studying DE models. While this is a specific
class of solutions, it serves as an excellent approximation for the
late-time universe. The qualitative insights derived from this model
such as the behavior of the EoS and stability analysis remain
relevant across various cosmological scenarios.

The deceleration parameter is defined as
\begin{equation}\label{19aa}
q=-\frac{a \ddot{a}}{\dot{a}^2} = -1 + \frac{1}{m},
\end{equation}
where dot means differentiation with respect to $t$. Substituting
the expression for $m$ into $a(t)$, we obtain
\begin{equation}\label{20aa}
a(t) = t^{\frac{1}{1+q}}.
\end{equation}
The deceleration parameter has a value of $ q =
-0.832^{+0.091}_{-0.091}$\cite{44a}, where $q > -1$ signifies an
expanding universe. This value also indicates the present
acceleration of the universe. Using this scale factor, the Hubble
parameter can be expressed as
\begin{equation}\label{21aa}
\emph{H}= (1+q)^{-1} t^{-1}, \quad \emph{H}_0 = (1+q)^{-1} t_0^{-1}.
\end{equation}
This implies that the expansion of the universe is governed by the
parameters $q $ and $\emph{H}_0$. The relationship between the
redshift parameter $z$ and the scale factor leads to
\begin{equation}\label{20}
\emph{H} = \emph{H}_0 \Upsilon^{1+q}, \quad \dot{\emph{H}} =
-\emph{H}_0 \Upsilon^{2+2q},\quad Q = 6\emph{H}_0^2 \Upsilon^{2+2q},
\quad \zeta=\frac{(q+1) \Upsilon^{-q}}{q}.
\end{equation}
where $\Upsilon = 1+z$. Integrating Eq.\eqref{17}, we obtain
\begin{equation}\label{21}
\varrho_{\mathbf{m}}=(a)^{-3} \xi,
\end{equation}
with $\xi$ being an integration constant. The modified Friedmann
equations are given by
\begin{equation}\label{22}
3\emph{H}^{2}=\varrho_D +\varrho_{\mathbf{m}},\quad
2\dot{\emph{H}}+3\emph{H}^{2}=\mathbb{P}_D +\mathbb{P}_{\mathbf{m}},
\end{equation}
with
\begin{eqnarray}\label{23}
\varrho_D&=&-6\emph{H}^{2}f_{Q}+\frac{f}{2},\\\label{24}
\mathbb{P}_D&=&2\emph{H}f_{QQ}+\frac{f}{2}+6\emph{H}^{2}f_{Q}+2f_{Q}\dot{\emph{H}}.
\end{eqnarray}

\section{Reconstruction of NADE $f(Q)$ Gravity Model}
\begin{figure}\center
\epsfig{file=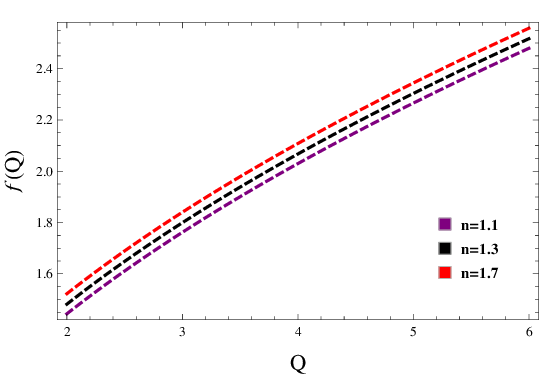,width=.5\linewidth}\caption{The graph of $f(Q)$
against $Q$.}
\end{figure}

In this section, we derive the relationship between NADE and the
$f(Q)$ gravity model by equating their corresponding densities. From
Eqs.\eqref{2} and \eqref{23}, we obtain
\begin{equation}\label{25}
-6\emph{H}^{2}f_{Q}+\frac{f}{2}=\frac{3n^{2}}{\zeta^{2}}.
\end{equation}
The solution to this first-order linear differential equation in $Q$
is given by
\begin{equation}\label{26}
f(Q)=\emph{c}\sqrt{Q}+\frac{12 n^2}{\zeta ^2}.
\end{equation}
To express this model in terms of $z$, we substitute Eq.\eqref{20}
into \eqref{26}, resulting in
\begin{equation}\label{27}
f(Q)=\sqrt{6} \emph{c} \sqrt{H_{0}^{2} (\Upsilon)^{2 q+2}}+\frac{12
n^2 q^2 (\Upsilon)^{2 q}}{(q+1)^2}.
\end{equation}
We set $\emph{c}=1$ for all our graphical analysis. We choose values
that yield well-behaved graphs and align with the properties of
NADE. The parametric graphs were constructed by selecting values
that satisfy the behavior of the two phase planes, as any changes in
these values directly impact the graphical behavior. Figures
\textbf{1} and \textbf{2} illustrate characteristics of the
reconstructed NADE $f(Q)$ gravity model at three various values of
$n$. The model shows an increasing trend with rising values of $Q$
and $z$, indicating that the NADE model represents accelerated
expansion.
\begin{figure}\center
\epsfig{file=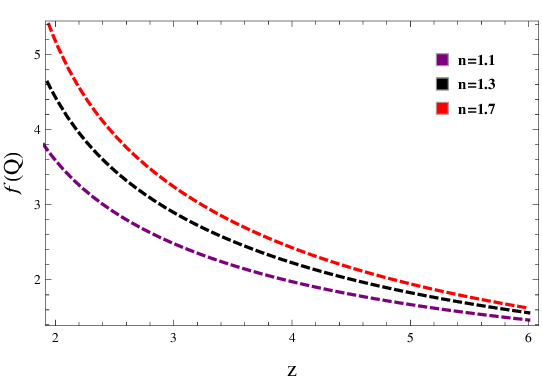,width=.5\linewidth}\caption{The graph
illustrates the connection between $f(Q)$ and $z$.}
\end{figure}

The values of $\varrho_D$ and $\mathbb{P}_D$ are found by
substituting Eq.\eqref{26} in \eqref{23} and \eqref{24} as
\begin{eqnarray}\label{a1}
\varrho_D&=&-\big(\sqrt{6} \emph{H}-\sqrt{Q}\big)\frac{1}{2}
\emph{c}+\frac{6 n^2}{\zeta ^2},\\\label{a2}
\mathbb{P}_D&=&\frac{\emph{c} \zeta ^2 \big(Q (2 \dot{\emph{H}}-Q)+6
\emph{H}^2 Q-\emph{H}\big)-12 n^2 Q^{3/2}}{2 \zeta ^2 Q^{3/2}}.
\end{eqnarray}
Using Eq.\eqref{20} in the above equations, we can express them as
\begin{eqnarray}\label{28}
\varrho_D&=&\sqrt{\frac{3}{2}} \emph{c} \bigg(\sqrt{\emph{H}_0^2
\Upsilon^{2 q+2}}-\emph{H}_0 \Upsilon^{q+1}\bigg)+\frac{6 n^2 q^2
\Upsilon^{2 q}}{(q+1)^2},
\\\nonumber \mathbb{P}_D&=&\bigg(q^2 \Upsilon^{2 q} \bigg(\frac{c \emph{H}_0 (q+1)^2 \Upsilon^{1-q}
 \big(-12 \emph{H}_0^2 \Upsilon^{3 q+3}-1\big)}{q^2}72 \sqrt{6}\\\label{29}&\times& n^2 \big(\emph{H}_0^2
 \Upsilon^{2 q+2}\big)^{3/2}\bigg)\bigg)\bigg(12 \sqrt{6} (q+1)^2 \big(\emph{H}_0^2 \Upsilon^{2 q+2}\big)^{3/2}\bigg)^{-1}.
\end{eqnarray}
The behavior of $\varrho_D$ and $\mathbb{P}_D$ over time is crucial
for understanding DE. Figure \textbf{3} shows that $\varrho_D$
increases as redshift increases, playing a key part in accelerating
the expansion of the cosmos. Figure \textbf{4} demonstrates that
$\mathbb{P}_D$ is decreasing, which is a defining characteristic of
DE. This decreasing pressure opposes gravitational forces, enabling
DE to push galaxies apart and fuel the ongoing expansion of the
cosmos.
\begin{figure}\center
\epsfig{file=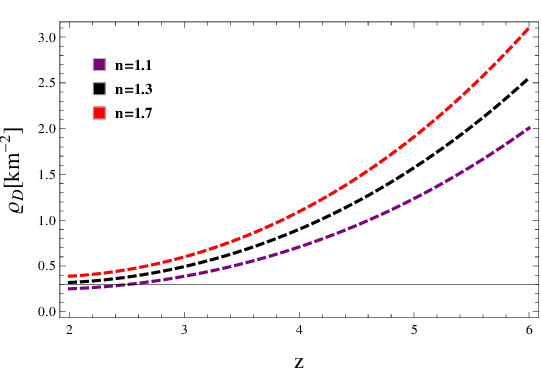,width=.5\linewidth}\caption{The graph of
$\varrho$ versus $z$.}
\end{figure}
\begin{figure}\center
\epsfig{file=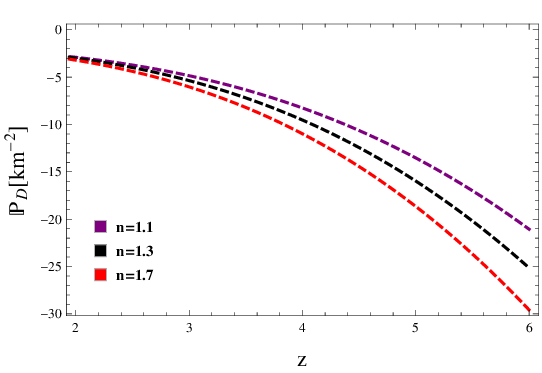,width=.5\linewidth}\caption{The plot of
$\mathbb{P}$ against $z$.}
\end{figure}

The equation of state (EoS) parameter ($\omega_{D}$) links
$\mathbb{P}_D$ to $\varrho_D$ using the formula
$\omega_{D}=\frac{\mathbb{P}_D}{\varrho_D}$. This parameter is
important for identifying different types of energy in the cosmos.
In the case of DE, when $\omega_{D}= -1$, it represents a vacuum. If
$\omega_{D}<-1$, it leads to a phantom phase, where the universe
expands even faster. The quintessence phase occurs when
$-1<\omega_{D}<-\frac{1}{3}$, which also affects the rate of cosmic
expansion. Thus we have
\begin{eqnarray}\nonumber
\omega_{D}&=&-\bigg(\frac{\emph{c} \emph{H}_0 (q+1)^2 \Upsilon^{1-q}
\big(12 \emph{H}_0^2 \Upsilon^{3 q+3}+1\big)}{q^2}+72 \sqrt{6} n^2
\big(\emph{H}_0^2 \Upsilon^{2 q+2}\big)^\frac{3}{2}\bigg)\bigg(6
\\\nonumber&\times&\sqrt{6}\bigg(\emph{H}_0^2
 \Upsilon^{2 q+2}\bigg)^\frac{3}{2} \bigg(\frac{(q+1)^2
\Upsilon^{-2 q}}{q^2}\bigg(\sqrt{6} c \bigg(\sqrt{\emph{H}_0^2
\Upsilon^{2 q+2}}-\emph{H}_0 \Upsilon^{q+1}\bigg)+2
\\\label{2a}&\times&\xi \Upsilon^3\bigg)+12 n^2\bigg)\bigg)^{-1}.
\end{eqnarray}
In Figure \textbf{5}, the EoS parametric value is observed within
the range $-1<\omega_{D}<-\frac{1}{3}$, confirming that the model
behaves like quintessence. This behavior suggests that the NADE
model effectively contributes to the rapid expansion of the cosmos
while remaining consistent with observational data.
\begin{figure}\center
\epsfig{file=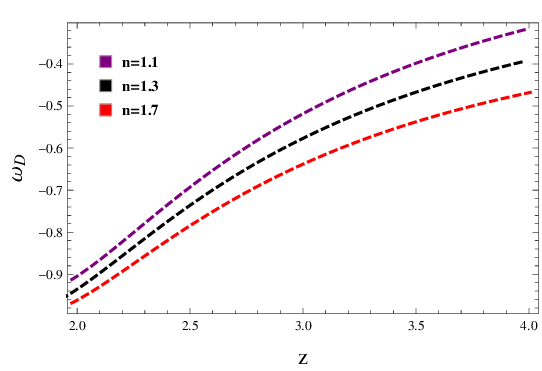,width=.5\linewidth}\caption{The graph shows the
relationship between $\omega_{D}$ and $z$.}
\end{figure}

Studying the phase plane defined by the
$\omega_{D}-\omega^{\prime}_{D}$ \cite{42} coordinates is crucial
for understanding the dynamic properties of DE models, where the
prime indicates a derivative with respect to the variable $Q$. This
phase plane provides significant insights into the evolution of DE
as the universe expands, revealing different phases of DE behavior
by categorizing the plane into two distinct regions based on the
characteristics of the EoS. The thawing region, where $\omega_{D}<0$
and $\omega^{\prime}_{D}>0$, suggests a period of less accelerated
expansion. In contrast, the freezing region, characterized by
$\omega_{D}<0$ and $\omega^{\prime}_{D}<0$, signifies a phase where
the expansion of the universe is more restrained. Here, we have
\begin{eqnarray}\nonumber
\omega^{\prime}_{D}&=&\bigg(\emph{c} (q+1)^2 \Upsilon^{-9 q-4}
\bigg(\emph{c} \emph{H}_0 (q+1)^2 \Upsilon^q \bigg(-4 \emph{H}_0
\Upsilon^{q+1} \bigg(3 \emph{H}_0 \big(\sqrt{\emph{H}_0^2
\Upsilon^{2 q+2}}
\\\nonumber&-&6 \emph{H}_0 \sqrt{\emph{H}_0^2 \Upsilon^{2 q+2}}+2 \emph{H}_0 (3
\emph{H}_0-1) \Upsilon^{q+1}\big)\Upsilon^{2 q+2}-1\bigg)-3
\sqrt{\emph{H}_0^2 \Upsilon^{2 q+2}}\bigg)\\\nonumber&+&\frac{
\sqrt{\emph{H}_0^2 \Upsilon^{2 q+2}}}{\Upsilon}\sqrt{6}\bigg(6 n^2
q^2 \Upsilon^{2 q} \big(1-4 \emph{H}_0^2 (3 \emph{H}_0-1)\Upsilon^{3
q+3}\big)-\xi (q+1)^2 \Upsilon^3 \\\nonumber&\times& \bigg(4
\emph{H}_0^2 (6 \emph{H}_0-1) \Upsilon^{3
q+3}-1\bigg)\bigg)\bigg)\bigg)\bigg(72 \emph{H}_0^5
q^4\bigg(\bigg((q+1)^2 \Upsilon^{-2 q} \bigg(\sqrt{6} \emph{c}
\\\label{3a}&\times&\bigg(\sqrt{\emph{H}_0^2 \Upsilon^{2 q+2}}-\emph{H}_0
\Upsilon^{q+1}\bigg)+2 \xi \Upsilon^3\bigg)\bigg)\frac{1}{q^2}+12 n^2\bigg)^2\bigg)^{-1}.
\end{eqnarray}
Figure \textbf{6} reveals that both $\omega_{D}<0$ and
$\omega^{\prime}_{D}<0$ across the three examined values of $n$,
which are parameters associated with the NADE model. This
observation is significant as it highlights the presence of a
freezing region within the framework of DE dynamics.
\begin{figure}\center
\epsfig{file=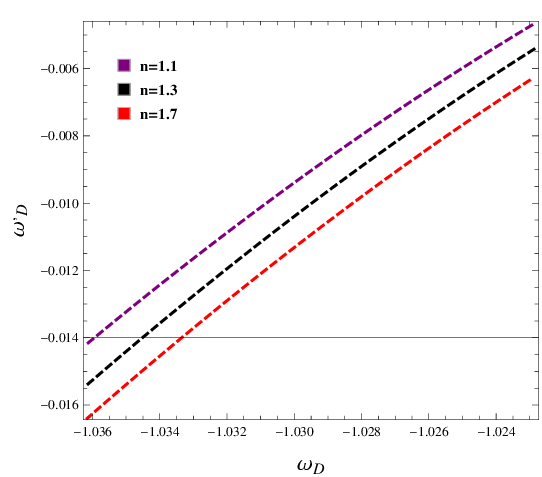,width=.5\linewidth}\caption{The graph
illustrates the connection between $\omega^{\prime}_{D}$ and
$\omega_{D}$.}
\end{figure}

The $(r-s)$-plane is useful for understanding how DE influences the
universe expansion \cite{43}. Different paths on this plane
correspond to different types of DE. When $r<1$ and $s>0$, the
trajectories fall within the quintessence and phantom phases. In
contrast, the Chaplygin gas model appears in the region where $r>1$
and $s<0$. The $r$ and $s$ are defined as
\begin{equation}\label{b1}
r=\frac{\dddot{a}}{aH^{3}}, \quad s=\frac{r-1}{3(q-\frac{1}{2})}.
\end{equation}
By applying Eq.\eqref{b1}, we obtain
\begin{eqnarray}\nonumber
r&=&\frac{1}{8} \bigg(16 \bigg(\bigg(\emph{c} (q+1)^2 \Upsilon^{-2
q-3} \bigg(\bigg(\emph{H}_0 \bigg(\bigg( \emph{c} (q+1)^2
\sqrt{\emph{H}_0^2 \Upsilon^{2 q+2}} \bigg( \big(\emph{H}_0^2
\Upsilon^{2 q+2}\big)^\frac{3}{2}\\\nonumber&\times&36-12 \emph{H}_0
\Upsilon^{2 q+2} \sqrt{\emph{H}_0^2 \Upsilon^{2 q+2}}+4\bigg)
\Upsilon^{-2 q}\sqrt{6}\bigg)\frac{1}{q^2}+36 n^2\bigg)
\Upsilon^{q+1}-3 \bigg(\bigg(\emph{c}
\\\nonumber&\times& (q+1)^2 \bigg(24 \sqrt{6} \bigg(\emph{H}_0^2
\Upsilon^{2 q+2}\bigg)^{3/2}+\sqrt{6}\bigg) \Upsilon^{-2
q}\bigg)\frac{1}{q^2}+144 \emph{H}_0^2 n^2 \Upsilon^{2
q+2}\bigg)\emph{H}_0^2
\\\nonumber&\times&\Upsilon^{2 q+2}+\bigg(36 \sqrt{6} \emph{c} \emph{H}_0^5 (q+1)^2
\Upsilon^{3 q+5}\bigg)\frac{1}{q^2}+24 \emph{H}_0^3 \bigg(\bigg(
(q+1)^2 \sqrt{\emph{H}_0^2 \Upsilon^{2 q+2}}
\\\nonumber&\times&\emph{c}\sqrt{6}\Upsilon^{-2 q}\bigg)\frac{1}{q^2}+6 n^2\bigg)
\Upsilon^{4 q+4}\bigg)\frac{1}{\Upsilon^3}-\bigg(2 (q+1)^2
\Upsilon^{-2 q} \bigg(-3 \emph{H}_0 \Upsilon^{q+1}+6
\\\nonumber&\times&\emph{H}_0^2 \bigg(6 \emph{H}_0^2 \Upsilon^{2 q+2}-2 \emph{H}_0
\Upsilon^{2 q+2}\bigg) \Upsilon^{2 q+2}+36 \emph{H}_0^4 \Upsilon^{4
q+4}\bigg)\xi\bigg)\frac{1}{q^2}\bigg)\bigg)\bigg(48\sqrt{6}q^2\\\nonumber&\times&
 \big(\emph{H}_0^2 \Upsilon^{2 q+2}\big)^\frac{5}{2}\big(\big(2 (q+1)^2 \xi
\Upsilon^{-2 q}\big)\frac{1}{q^2}+\big(\big(\emph{c} (q+1)^2
\big(\sqrt{6} \sqrt{\emph{H}_0^2 \Upsilon^{2 q+2}}-
\emph{H}_0\\\nonumber&\times&\sqrt{6} \Upsilon^{q+1}\big)
\Upsilon^{-2 q}\big)\frac{1}{q^2}+12
n^2\big)\frac{1}{\Upsilon^3}\big)^2\bigg)^{-1}+\frac{1}{2}\bigg)^2+\bigg(\emph{c}
(q+1)^2 \Upsilon^{q-3}\bigg(\bigg(\emph{H}_0 \\\nonumber&\times&
\bigg(\bigg(\sqrt{6} \emph{c} (q+1)^2 \sqrt{\emph{H}_0^2 \Upsilon^{2
q+2}} \bigg(36 \bigg(\emph{H}_0^2 \Upsilon^{2
q+2}\bigg)^{3/2}-\Upsilon^{2 q+2}\sqrt{\emph{H}_0^2 \Upsilon^{2
q+2}}\\\nonumber&\times& 12 \emph{H}_0  +4\bigg) \Upsilon^{-2
q}\bigg)\frac{1}{q^2}+36 n^2\bigg) \Upsilon^{q+1}-3 \emph{H}_0^2
\bigg(\big(\emph{c} (q+1)^2 \bigg(\big(\emph{H}_0^2 \Upsilon^{2
q+2}\big)^\frac{3}{2}
\\\nonumber&\times&24 \sqrt{6}+\sqrt{6}\bigg) \Upsilon^{-2
q}\big)\frac{1}{q^2}+144 \emph{H}_0^2 n^2 \Upsilon^{2 q+2}\bigg)
\Upsilon^{2 q+2}+\frac{1}{q^2} \emph{H}_0^5 (q+1)^2 36
\emph{c}\\\nonumber&\times&\sqrt{6}\Upsilon^{3 q+5}+24
\bigg(\frac{\sqrt{6} \emph{c} (q+1)^2 \sqrt{\emph{H}_0^2 \Upsilon^{2
q+2}} \Upsilon^{-2 q}}{q^2}+6 n^2\bigg)\Upsilon^{4 q+4}\emph{H}_0^3
\bigg)\frac{1}{\Upsilon^3}
\\\nonumber&-&\bigg(2(q+1)^2 \Upsilon^{-2 q} \bigg(-3 \emph{H}_0
\Upsilon^{q+1}+6 \emph{H}_0^2 \big(6 \emph{H}_0^2 \Upsilon^{2 q+2}-2
\emph{H}_0\Upsilon^{2 q+2}\big) \Upsilon^{2 q+2}\\\nonumber&+& 36
\emph{H}_0^4 \Upsilon^{4 q+4}\bigg)\xi
\bigg)\frac{1}{q^2}\bigg)\bigg)\bigg(6 \sqrt{6} q^2
\bigg(\emph{H}_0^2 \Upsilon^{2
q+2}\bigg)^{5/2}\bigg(\frac{1}{q^2}\big(2 (q+1)^2 \xi \Upsilon^{-2
q}\big)
\\\nonumber&+&\frac{1}{\Upsilon^3}\bigg(\big(\emph{c} (q+1)^2
\big(\sqrt{6} \sqrt{\emph{H}_0^2 \Upsilon^{2 q+2}}-\sqrt{6}
\emph{H}_0 \Upsilon^{q+1}\big)\Upsilon^{2 q}\big)\frac{1}{q^2}+12
n^2\bigg)\bigg)^2\bigg)^{-1}\\\nonumber&+& \bigg(\emph{c} (q+1)^2
\Upsilon^{-3 q-4} \bigg(\frac{1}{q^4}\big(4 (q+1)^4 \bigg(36
\emph{H}_0^2 \big( \emph{H}_0^2 \Upsilon^{2 q+2}- \emph{H}_0
\Upsilon^{2 q+2}\big) \Upsilon^{2 q+2}\\\nonumber&\times&-15
\emph{H}_0 \Upsilon^{q+1}+108 \emph{H}_0^4 \Upsilon^{4 q+4}\bigg)
\xi ^2 \Upsilon^{-4 q}\big)-\bigg(3 \emph{H}_0
\bigg(\frac{1}{q^2}\big(4 \sqrt{6} \emph{c} n^2 (q+1)^2
\\\nonumber&\times&\sqrt{\emph{H}_0^2 \Upsilon^{2 q+2}} \bigg(36 \bigg(\emph{H}_0^2 \Upsilon^{2
q+2}\bigg)^{3/2}+72 \emph{H}_0 \Upsilon^{2 q+2} \sqrt{\emph{H}_0^2
\Upsilon^{2 q+2}}+37\bigg) \Upsilon^{-2
q}\big)\\\nonumber&+&\frac{1}{q^4}\big(6 \emph{c}^2 \emph{H}_0^2
(q+1)^4 \bigg(36 \bigg(\emph{H}_0^2 \Upsilon^{2 q+2}\bigg)^{3/2}-36
\emph{H}_0 \Upsilon^{2 q+2} \sqrt{\emph{H}_0^2 \Upsilon^{2
q+2}}+8\bigg)\\\nonumber&\times& \Upsilon^{2-2 q}\big)+720 n^4\bigg)
\Upsilon^{q+1}-\frac{1}{q^2}\big(18 \emph{c} \emph{H}_0^3 (q+1)^2
\bigg(\frac{1}{q^2}\big(\emph{c} (q+1)^2 \bigg(5+108
\\\nonumber&\times&\bigg(\emph{H}_0^2 \Upsilon^{2 q+2}\bigg)^{3/2}\bigg)
\Upsilon^{-2 q}\big)+144 \sqrt{6} \emph{H}_0^2 n^2 \Upsilon^{2
q+2}\bigg) \Upsilon^{q+3}\big)+\emph{H}_0^2
\bigg(\frac{1}{q^4}\big(\sqrt{6}
\\\nonumber&\times&\emph{c}^2 (q+1)^4 \sqrt{\emph{H}_0^2 \Upsilon^{2 q+2}} \bigg(324
\sqrt{6} \bigg(\emph{H}_0^2 \Upsilon^{2 q+2}\bigg)^{3/2}+37
\sqrt{6}- \sqrt{6} \emph{H}_0 \Upsilon^{2
q+2}\\\nonumber&\times&36\sqrt{\emph{H}_0^2 \Upsilon^{2 q+2}}\bigg)
\Upsilon^{-4 q}\big)+\frac{1}{q^2}\big(72 \emph{c} n^2 (q+1)^2 \big(
\sqrt{6} \big(\emph{H}_0^2 \Upsilon^{2 q+2}\big)^{3/2}+5
\sqrt{6}\big)\\\nonumber&\times& \Upsilon^{-2 q}\big)+15552
\emph{H}_0^2 n^4 \Upsilon^{2 q+2}\bigg) \Upsilon^{2 q+2}+\frac{648
\emph{c}^2 \emph{H}_0^6 (q+1)^4 \Upsilon^{2q+6}}{q^4}-48
\emph{H}_0^3\\\nonumber&\times& \bigg(\frac{1}{q^2}\big(27 \sqrt{6}
\emph{c} n^2 (q+1)^2 \sqrt{\emph{H}_0^2 \Upsilon^{2 q+2}}
\Upsilon^{-2 q}\big)+\frac{\emph{c}^2 \emph{H}_0^2 (q+1)^4
\Upsilon^{2-2 q}}{q^4}+108\\\nonumber&\times&  n^4\bigg) \Upsilon^{4
q+4}\bigg)\frac{1}{\Upsilon^6}+\frac{1}{q^2}\big(2 \Upsilon^{-2 q-3}
\bigg(\emph{H}_0 \bigg(\frac{1}{q^2}\big(\sqrt{6} \emph{c} (q+1)^2
\sqrt{\emph{H}_0^2 \Upsilon^{2 q+2}} \bigg(72
\\\nonumber&\times&\big(\emph{H}_0^2 \Upsilon^{2 q+2}\big)^{3/2}-72
\emph{H}_0 \Upsilon^{2 q+2} \sqrt{\emph{H}_0^2 \Upsilon^{2
q+2}}+37\bigg) \Upsilon^{2 q}\big)+360 n^2\bigg)
\Upsilon^{q+1}\\\nonumber&\times&(q+1)^2+6 \emph{H}_0^2
\bigg(\frac{1}{q^2}\big(\emph{c} (q+1)^2 \bigg(54 \sqrt{6}
\bigg(\emph{H}_0^2 \Upsilon^{2 q+2}\bigg)^{3/2}+5 \sqrt{6}\bigg)
\Upsilon^{-2 q}\big)\\\nonumber&+&432 \emph{H}_0^2 n^2 \Upsilon^{2
q+2}\bigg) \Upsilon^{2 q+2}+18 \emph{H}_0^2
\bigg(\frac{1}{q^2}\big(6 \sqrt{6} \emph{c} (q+1)^2 \Upsilon^{-2 q}
\big(\emph{H}_0^2 \Upsilon^{2
q+2}\big)^\frac{3}{2}\big)\\\nonumber&+&24 \emph{H}_0^2 n^2
\Upsilon^{2 q+2}-6 \emph{H}_0 \Upsilon^{2 q+2} \bigg(\frac{\sqrt{6}
\emph{c} (q+1)^2 \sqrt{\emph{H}_0^2 \Upsilon^{2 q+2}} \Upsilon^{-2
q}}{q^2}+8 n^2\bigg)\bigg)
\\\nonumber&\times&\Upsilon^{2 q+2}-\frac{216 \sqrt{6} \emph{c}
\emph{H}_0^5 (q+1)^2 \Upsilon^{3 q+5}}{q^2}\bigg)
\xi\big)\bigg)\bigg)\bigg(72 \sqrt{6} \emph{H}_0 q^2
\bigg(\emph{H}_0^2 \Upsilon^{2 q+2}\bigg)^\frac{7}{2}
\\\nonumber&\times&\bigg(\frac{1}{\Upsilon^3}\big(\frac{1}{q^2}\big(\emph{c} (q+1)^2 \Upsilon^{-2
q} \bigg(\sqrt{6} \emph{H}_0 \Upsilon^{q+1}-\sqrt{6}
\sqrt{\emph{H}_0^2 \Upsilon^{2 q+2}}\bigg)\big)-12
n^2\big)\\\label{4a}&-&\frac{2 (q+1)^2 \Upsilon^{-2 q} \xi
}{q^2}\bigg)^3\bigg)^{-1}+4\bigg),
\\\nonumber
s&=& \bigg(2 \sqrt{6} q^2 \Upsilon^{2 q+3} \big(\emph{H}_0^2
\Upsilon^{2 q+2}\big)^\frac{5}{2} \bigg(\frac{2 (q+1)^2 \xi
\Upsilon^{-2
q}}{q^2}+\frac{1}{\Upsilon^3}\big(\frac{1}{q^2}\big(\emph{c} (q+1)^2
\big(\sqrt{6}\\\nonumber&\times& \sqrt{\emph{H}_0^2 \Upsilon^{2
q+2}}-\sqrt{6} \emph{H}_0 \Upsilon^{q+1}\big) \Upsilon^{-2
q}\big)+12 n^2\big)\bigg)^2 \bigg(16 \bigg(\bigg(\emph{c} (q+1)^2
\Upsilon^{2 q-3}\\\nonumber&\times&
\bigg(\bigg(\emph{H}_0\bigg(\frac{1}{q^2}\big(\sqrt{6} \emph{c}
(q+1)^2 \sqrt{\emph{H}_0^2 \Upsilon^{2 q+2}} \bigg(36
\big(\emph{H}_0^2 \Upsilon^{2 q+2}\big)^{3/2}-12 \emph{H}_0
\Upsilon^{2 q+2}\\\nonumber&\times& \sqrt{\emph{H}_0^2 \Upsilon^{2
q+2}}+4\bigg) \Upsilon^{-2 q}\big)+36 n^2\bigg) \Upsilon^{q+1}-3
\emph{H}_0^2 \bigg(\frac{1}{q^2}\big(\emph{c} (q+1)^2 \bigg(24
\sqrt{6}\\\nonumber&\times& \bigg(\emph{H}_0^2 \Upsilon^{2
q+2}\bigg)^{3/2}+\sqrt{6}\bigg) \Upsilon^{-2 q}\big)+144
\emph{H}_0^2 n^2 \Upsilon^{2 q+2}\bigg) \Upsilon^{2 q+2}+36 \sqrt{6}
\emph{c} \emph{H}_0^5\\\nonumber&\times&\frac{(q+1)^2 \Upsilon^{3
q+5}}{q^2}+24 \emph{H}_0^3 \bigg(\frac{\sqrt{6} \emph{c} (q+1)^2
\sqrt{\emph{H}_0^2 \Upsilon^{2 q+2}} \Upsilon^{-2 q}}{q^2}+6
n^2\bigg)
\Upsilon^{4q+4}\bigg)\\\nonumber&\times&\frac{1}{\Upsilon^3} \big(2
(q+1)^2 \Upsilon^{2 q} \bigg(3 \emph{H}_0 \Upsilon^{q+1}+6
\emph{H}_0^2 \bigg(6 \emph{H}_0^2 \Upsilon^{2 q+2}-2 \emph{H}_0
\Upsilon^{2 q+2}\bigg) \Upsilon^{2 q+2}\\\nonumber&+&36 \emph{H}_0^4
\Upsilon^{4 q+4}\bigg) \xi\big) -\frac{1}{q^2}\bigg)\bigg)\bigg(48
\sqrt{6} q^2 \bigg(\emph{H}_0^2 \Upsilon^{2 q+2}\bigg)^{5/2}
\bigg(\frac{2 (q+1)^2 \xi \Upsilon^{-2
q}}{q^2}\\\nonumber&+&\frac{1}{\Upsilon^3}\big(\frac{1}{q^2}\big(\emph{c}
(q+1)^2 \bigg(\sqrt{6} \sqrt{\emph{H}_0^2 \Upsilon^{2 q+2}}-\sqrt{6}
\emph{H}_0 \Upsilon^{q+1}\bigg) \Upsilon^{-2 q}\big)+12
n^2\big)\bigg)^2\bigg)^{-1}\\\nonumber&+&\frac{1}{2}\bigg)^2+\bigg(\emph{c}
(q+1)^2 \Upsilon^{-2 q-3} \bigg(\bigg(\emph{H}_0
\bigg(\frac{1}{q^2}\big(\sqrt{6} \emph{c} (q+1)^2 \sqrt{\emph{H}_0^2
\Upsilon^{2 q+2}} \big(36\big(\emph{H}_0^2 \\\nonumber&\times&
\Upsilon^{2 q+2}\big)^{3/2}-12 \emph{H}_0 \Upsilon^{2 q+2}
\sqrt{\emph{H}_0^2 \Upsilon^{2 q+2}}+4\big) \Upsilon^{-2 q}\big)+36
n^2\bigg) \Upsilon^{q+1}-3 \emph{H}_0^2\\\nonumber&\times&
\bigg(\frac{1}{q^2}\big(\emph{c} (q+1)^2 \big(24 \sqrt{6}
\big(\emph{H}_0^2 \Upsilon^{2 q+2}\big)^{3/2}+\sqrt{6}\big)
\Upsilon^{-2 q}\big)+144 \emph{H}_0^2 n^2 \Upsilon^{2 q+2}\bigg)
\\\nonumber&\times&\Upsilon^{2 q+2}+\frac{36 \sqrt{6} \emph{c} \emph{H}_0^5 (q+1)^2
\Upsilon^{3 q+5}}{q^2}+ \bigg(\frac{\sqrt{6} \emph{c} (q+1)^2
\sqrt{\emph{H}_0^2 \Upsilon^{2 q+2}} \Upsilon^{-2
q}}{q^2}\\\nonumber&+&6 n^2\bigg)24 \emph{H}_0^3 \Upsilon^{4
q+4}\bigg)\frac{1}{\Upsilon^3}-\frac{1}{q^2}\big(2 (q+1)^2
\Upsilon^{-2 q} \bigg(-3 \emph{H}_0 \Upsilon^{q+1}+6
\emph{H}_0^2\big(6\\\nonumber&\times&  \emph{H}_0^2 \Upsilon^{2
q+2}-2 \emph{H}_0 \Upsilon^{2 q+2}\big) \Upsilon^{2 q+2}+36
\emph{H}_0^4 \Upsilon^{4 q+4}\bigg) \xi \big)\bigg)\bigg)\bigg(6
\big(\emph{H}_0^2 \Upsilon^{2 q+2}\big)^{5/2}\\\nonumber&\times&
\sqrt{6} q^2\bigg(\frac{2 (q+1)^2 \xi \Upsilon^{-2
q}}{q^2}+\frac{1}{\Upsilon^3}\big(\big(\emph{c} (q+1)^2
\bigg(\sqrt{6} \sqrt{\emph{H}_0^2 \Upsilon^{2 q+2}}-\sqrt{6}
\emph{H}_0
\\\nonumber&\times&\Upsilon^{q+1}\bigg)\Upsilon^{-2
q}\big)\frac{1}{q^2}+12 n^2\big)\bigg)^2\bigg)^{-1}+\bigg(\emph{c}
(q+1)^2 \Upsilon^{-3 q-4} \bigg(\frac{1}{q^4}\big(4 (q+1)^4
\\\nonumber&\times&\bigg(108 \emph{H}_0^4
\Upsilon^{4 q+4}-15 \emph{H}_0 \Upsilon^{q+1}+6 \emph{H}_0^2 \big(6
\emph{H}_0^2 \Upsilon^{2 q+2}-6 \emph{H}_0 \Upsilon^{2 q+2}\big)
\Upsilon^{2 q+2}\bigg)\\\nonumber&\times&\xi ^2 \Upsilon^{-4
q}\big)+\bigg(3 \emph{H}_0 \bigg(\frac{1}{q^2}\big(4 \sqrt{6}
\emph{c} n^2 (q+1)^2 \sqrt{\emph{H}_0^2 \Upsilon^{2 q+2}} \bigg(36
\big(\emph{H}_0^2 \Upsilon^{2 q+2}\big)^{3/2}\\\nonumber&-&72
\emph{H}_0 \Upsilon^{2 q+2} \sqrt{\emph{H}_0^2 \Upsilon^{2
q+2}}+37\bigg) \Upsilon^{-2 q}\big)+\frac{1}{q^4}\big(6 \emph{c}^2
\emph{H}_0^2 (q+1)^4 \bigg(36 \big( \Upsilon^{2
q+2}\\\nonumber&\times&\emph{H}_0^2\big)^{3/2}-36 \emph{H}_0
\Upsilon^{2 q+2} \sqrt{\emph{H}_0^2 \Upsilon^{2 q+2}}+8\bigg)
\Upsilon^{2-2 q}\big)+720 n^4\bigg)
\Upsilon^{q+1}-18\\\nonumber&\times&\frac{1}{q^2}\big(\emph{c}
\emph{H}_0^3 (q+1)^2 \bigg(\frac{1}{q^2}\big(\emph{c} (q+1)^2
\bigg(108 \big(\emph{H}_0^2 \Upsilon^{2 q+2}\big)^{3/2}+5\bigg)
\Upsilon^{-2 q}\big)+ \sqrt{6}\\\nonumber&\times&144 \emph{H}_0^2
n^2 \Upsilon^{2 q+2}\bigg ) \Upsilon^{q+3}\big)+\emph{H}_0^2
\bigg(\frac{1}{q^4}\big(\sqrt{6} \emph{c}^2 (q+1)^4
\sqrt{\emph{H}_0^2 \Upsilon^{2 q+2}} \bigg(324
\sqrt{6}\\\nonumber&\times& \big(\emph{H}_0^2 \Upsilon^{2
q+2}\big)^{3/2}-36 \sqrt{6} \emph{H}_0 \Upsilon^{2 q+2}
\sqrt{\emph{H}_0^2 \Upsilon^{2 q+2}}+37 \sqrt{6}\bigg) \Upsilon^{-4
q}\big)+\frac{1}{q^2}\big(72 \\\nonumber&\times&\emph{c} n^2 (q+1)^2
\big(54 \sqrt{6} \big(\emph{H}_0^2 \Upsilon^{2 q+2}\big)^{3/2}+5
\sqrt{6}\big) \Upsilon^{-2 q}\big)+15552 \emph{H}_0^2 n^4
\Upsilon^{2 q+2}\bigg)\\\nonumber&\times& \Upsilon^{2 q+2}+\frac{648
\emph{c}^2 \emph{H}_0^6 (q+1)^4 \Upsilon^{2 q+6}}{q^4}-48
\emph{H}_0^3 \bigg(\frac{27 \sqrt{6} \emph{c}  \sqrt{\emph{H}_0^2
\Upsilon^{2 q+2}} \Upsilon^{-2 q}}{q^2}\\\nonumber&\times&n^2
(q+1)^2+\frac{12 \emph{c}^2 \emph{H}_0^2 (q+1)^4 \Upsilon^{2-2
q}}{q^4}+108 n^4\bigg) \Upsilon^{4 q+4}\bigg)\frac{1}{\Upsilon^6}+2
(q+1)^2\\\nonumber&\times&\bigg( \Upsilon^{-2 q-3} \bigg(-\emph{H}_0
\bigg(\frac{1}{q^2}\big(\sqrt{6} \emph{c} (q+1)^2 \sqrt{\emph{H}_0^2
\Upsilon^{2 q+2}} \big(72 \big(\emph{H}_0^2 \Upsilon^{2
q+2}\big)^{3/2}-72\\\nonumber&\times& \emph{H}_0 \Upsilon^{2 q+2}
\sqrt{\emph{H}_0^2 \Upsilon^{2 q+2}}+37\big) \Upsilon^{-2
q}\big)+360 n^2\bigg) \Upsilon^{q+1}+6 \emph{H}_0^2
\bigg(\big(\emph{c} (q+1)^2
\\\nonumber&\times&\big(54 \sqrt{6} \big(\emph{H}_0^2 \Upsilon^{2
q+2}\big)^{3/2}+5 \sqrt{6}\big) \Upsilon^{-2
q}\big)\frac{1}{q^2}+432 \emph{H}_0^2 n^2 \Upsilon^{2 q+2}\bigg)
\Upsilon^{2 q+2}+8\\\nonumber&\times& \emph{H}_0^2
\bigg(\frac{1}{q^2}\big(6 \sqrt{6} \emph{c} (q+1)^2 \Upsilon^{-2 q}
\bigg(\emph{H}_0^2 \Upsilon^{2 q+2}\bigg)^{3/2}\big)+24 \emph{H}_0^2
n^2 \Upsilon^{2 q+2}-6 \emph{H}_0\\\nonumber&\times&\Upsilon^{2 q+2}
\bigg(\frac{\sqrt{6} \emph{c} (q+1)^2 \sqrt{\emph{H}_0^2 \Upsilon^{2
q+2}} \Upsilon^{-2 q}}{q^2}+8 n^2\bigg)\bigg) \Upsilon^{2 q+2}-216
\sqrt{6} \emph{c} \emph{H}_0^5\\\nonumber&\times&\frac{(q+1)^2
\Upsilon^{3 q+5}}{q^2}\bigg) \xi
\bigg)\frac{1}{q^2}\bigg)\bigg)\bigg(72 \sqrt{6} \emph{H}_0 q^2
\big(\emph{H}_0^2 \Upsilon^{2 q+2}\big)^{7/2}
\bigg(\frac{1}{\Upsilon^3}\big(\frac{1}{q^2}\big(\emph{c}\Upsilon^{-2
q} \\\nonumber&\times&(q+1)^2 \sqrt{6}\big( \emph{H}_0
\Upsilon^{q+1}- \sqrt{\emph{H}_0^2 \Upsilon^{2 q+2}}\big)\big)-12
n^2\big)-\frac{2 (q+1)^2 \Upsilon^{-2 q} \xi
}{q^2}\bigg)^3\bigg)^{-1}\\\nonumber&-&4\bigg)\bigg)\bigg(\emph{c}
(q+1)^2 \bigg(\bigg(\emph{H}_0 \bigg(\frac{1}{q^2}\big(\sqrt{6}
\emph{c} (q+1)^2 \sqrt{\emph{H}_0^2 \Upsilon^{2 q+2}} \bigg(36
\bigg(\emph{H}_0^2 \Upsilon^{2 q+2}\bigg)^{3/2}\\\nonumber&-&12
\emph{H}_0 \Upsilon^{2 q+2} \sqrt{\emph{H}_0^2 \Upsilon^{2
q+2}}+4\bigg) \Upsilon^{-2 q}\big)+36 n^2\bigg) \Upsilon^{q+1}-3
\emph{H}_0^2 \emph{c}
(q+1)^2\\\nonumber&\times&\bigg(\frac{1}{q^2}\big(\big(24 \sqrt{6}
\bigg(\emph{H}_0^2 \Upsilon^{2 q+2}\bigg)^{3/2}+\sqrt{6}\big)
\Upsilon^{-2 q}\big)+144 \emph{H}_0^2 n^2 \Upsilon^{2 q+2}\bigg)
\Upsilon^{2 q+2}\\\nonumber&+&\frac{36 \sqrt{6} \emph{c}
\emph{H}_0^5 (q+1)^2 \Upsilon^{3 q+5}}{q^2}+24 \emph{H}_0^3
\bigg(\frac{\sqrt{6} \emph{c} (q+1)^2 \sqrt{\emph{H}_0^2 \Upsilon^{2
q+2}} \Upsilon^{-2 q}}{q^2}+6\\\nonumber&\times&n^2\bigg)
\Upsilon^{4 q+4}\bigg)\frac{1}{\Upsilon^3}-\frac{1}{q^2}\big(2
(q+1)^2 \Upsilon^{-2 q} \bigg(-3 \emph{H}_0 \Upsilon^{q+1}+6
\emph{H}_0^2 \bigg(6 \emph{H}_0^2 \Upsilon^{2 q+2}\\\label{4b}&-&2
\emph{H}_0 \Upsilon^{2 q+2}\bigg) \Upsilon^{2 q+2}+36 \emph{H}_0^4
\Upsilon^{4 q+4}\bigg) \xi \big)\bigg)\bigg)^{-1}.
\end{eqnarray}
Figure \textbf{7} illustrates the $(r-s)$-plane with different
values of $n$ in the context of the NADE framework. The figure
specifically highlights the regions where $r$ is greater than 1 and
$s$ is less than 0, indicating the characteristics and consequences
of the Chaplygin gas model.
\begin{figure}\center
\epsfig{file=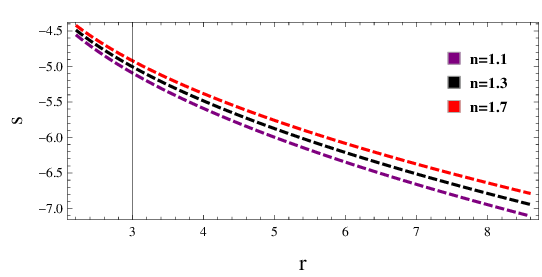,width=.5\linewidth}\caption{The graph
illustrates the connection between $r$ and $s$.}
\end{figure}

The squared speed of sound ($\nu_{s}^{2}$) in a fluid or
cosmological context is a crucial parameter that helps to understand
the stability and behavior of DE models. A positive value of
$\nu_{s}^{2}$ indicates stability, while a negative value signals
instability. Recognizing these characteristics is essential for
evaluating the viability of different models and their implications
for the universe expansion. This is given as
\begin{equation}\nonumber
\nu_{s}^{2}=\frac{\dot{p}_{D}}{\dot{\varrho}_{D}}
=\frac{\varrho_D}{{\dot{\varrho}_{D}}}\omega^{\prime}_{D}+\omega_{D},
\end{equation}
and hence
\begin{eqnarray}\nonumber
\nu_{s}^{2}&=&\bigg(-2 \bigg(\frac{\emph{c} \emph{H}_0 (q+1)^2
\Upsilon^{1-q} \big(12 \emph{H}_0^2 \Upsilon^{3 q+3}+1\big)}{q^2}+72
\sqrt{6} n^2 \big(\emph{H}_0^2 \Upsilon^{2
q+2}\big)^{3/2}\bigg)\\\nonumber&\times&
\bigg(\big(\frac{1}{q^2}\big(\sqrt{6} \emph{c} (q+1)^2 \Upsilon^{-2
q} \big(\sqrt{\emph{H}_0^2 \Upsilon^{2 q+2}}-\emph{H}_0
\Upsilon^{q+1}\big)\big)+12 n^2\big)\frac{1}{\Upsilon^3}+2 \xi
\\\nonumber&\times&\frac{(q+1)^2 \Upsilon^{-2 q}}{q^2}\bigg)-\bigg(\emph{c} (q+1)^2
\Upsilon^{-7 q-5} \big(24 \emph{H}_0^3 \Upsilon^{3 q+3}-4
\emph{H}_0^2 \Upsilon^{3
q+3}-1\big)\\\nonumber&\times&\bigg(\emph{H}_0 \Upsilon^{q-2}
\bigg(\frac{36 \sqrt{6} \emph{c} \emph{H}_0^4 (q+1)^2 \Upsilon^{2
q+4}}{q^2}-3 \emph{H}_0 \Upsilon^{q+1} \bigg(\sqrt{6}
\emph{c}(q+1)^2 \Upsilon^{-2 q}\\\nonumber&\times&\frac{ \big(24
\big(\emph{H}_0^2 \Upsilon^{2 q+2}\big)^{3/2}+1\big)}{q^2}+144
\emph{H}_0^2 n^2 \Upsilon^{2 q+2}\bigg)+24 \emph{H}_0^2 \Upsilon^{3
q+3} \bigg((q+1)^2\\\nonumber&\times&\frac{\sqrt{6} \emph{c}
 \Upsilon^{-2 q} \sqrt{\emph{H}_0^2 \Upsilon^{2 q+2}}}{q^2}+6
n^2\bigg)+\frac{1}{q^2}\big(\sqrt{6} \emph{c} (q+1)^2 \Upsilon^{-2
q} \bigg(\big(\emph{H}_0^2 \Upsilon^{2
q+2}\big)^{3/2}\\\nonumber&\times&36+4-\frac{ \big(12\emph{H}_0^2
\Upsilon^{2 q+2}\big)^{3/2}}{\emph{H}_0}\bigg) \sqrt{\emph{H}_0^2
\Upsilon^{2 q+2}}\big)+36 n^2\bigg)-6 \emph{H}_0 \xi (q+1)^2
\\\nonumber&\times&\frac{1}{q^2}\big(\Upsilon^{1-q} \bigg(24 \emph{H}_0^3
\Upsilon^{3 q+3}-4 \emph{H}_0^2 \Upsilon^{3
q+3}-1\bigg)\big)\bigg)\bigg)\frac{1}{72 \emph{H}_0^5
q^2}\bigg)\bigg(12 \sqrt{6} \Upsilon^3
\\\nonumber&\times&\big(\emph{H}_0^2 \Upsilon^{2 q+2}\big)^{3/2}
\bigg(\frac{1}{\Upsilon^3}\big(\frac{1}{q^2}\big(\sqrt{6} \emph{c}
(q+1)^2 \Upsilon^{-2 q} \big(\sqrt{\emph{H}_0^2 \Upsilon^{2
q+2}}-\emph{H}_0 \Upsilon^{q+1}\big)\big)\\\label{6aa}&+&12
n^2\big)+\frac{2 \xi (q+1)^2 \Upsilon^{-2
q}}{q^2}\bigg)^2\bigg)^{-1}.
\end{eqnarray}
The NADE model consistently shows negative values, signifying the
model's instability. Several studies have highlighted this
instability in the NADE model \cite{44}. Figure \textbf{8}
illustrates the behavior of $\nu_{s}^{2}$, which is observed to be
negative across different values of $n$ with NADE model. These
negative values indicate that the system remains unstable as the
universe evolves. This aligns with earlier research, indicating that
the $f(Q)$ model encounters similar instability issues.
\begin{figure}\center
\epsfig{file=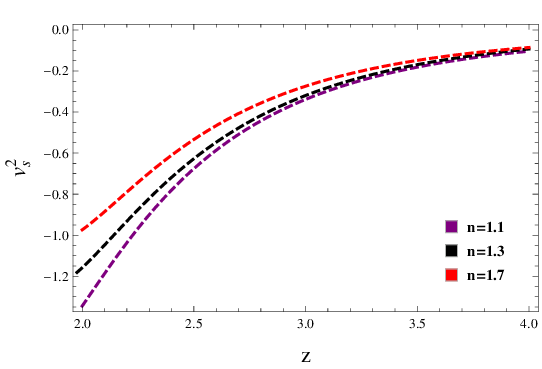,width=.5\linewidth}\caption{The graph depicts
the relationship between $\nu_{s}^{2}$ and $z$.}
\end{figure}

\section{Conclusions}

In this manuscript, we have explored the NADE model within the
$f(Q)$ gravity, focusing on a non-interacting scenario. We have
reconstructed the NADE model by applying a correspondence method to
the FRW model and assumed a power-law solution to analyze cosmic
expansion. Our investigation has covered the universe evolution
through the EoS, $\omega_{D}-\omega^{\prime}_{D}$ and $(r-s)$ planes
and examined the stability of the model using $\nu_{s}^{2}$. The key
findings of this study are summarized below.
\begin{itemize}
\item The NADE $f(Q)$ gravity model exhibits an increase
with respect to $z$ and $Q$. This suggests that the reconstructed
model is credible. (Figures \textbf{1} and \textbf{2}).
\item For all values of $n$, the $\varrho_D$ increases
while the $\mathbb{P}_D$ decreases in NADE model. This indicates
that the model contributes to the rapid expansion observed today.
Its alignment with fundamental DE properties reinforces their
validity as an explanation for one of the key phenomena in cosmology
(Figures \textbf{3} and \textbf{4}).
\item To characterize various cosmic epochs, we have plotted the EoS parameter
against time $z$ for differen $n$ values. In these cases,
$\omega_{D}$ shows quintessence-like behavior for the power-law
form. It is observed that the current rate of change in energy
density may be slow enough to address the coincidence problem
(Figure \textbf{5}).
\item The $(\omega_{D}-\omega^{\prime}_{D})$-plane represents a freezing
region, indicating that the universe appears to be expanding at an
accelerated rate (Figure \textbf{6}).
\item We have observed that the behavior of the $(r-s)$-plane
illustrates the Chaplygin gas model (Figure \textbf{7}).
\item We have examined the model stability through $\nu_{s}^{2}$ and
observed that the NADE $f(Q)$ model remains unstable throughout the
entire development of cosmos (Figure \textbf{8}).
\end{itemize}

We have conducted a detailed comparison of our model with key
observational constraints including CMB, SNeIa and BAO. We outline
how our model aligns with or deviates from these observational
datasets. The CMB provides precise constraints on the DE EoS
$(\omega_{D})$. Recent Planck 2018 results give the following
estimates at a 85$\%$ confidence level.
\begin{eqnarray}\nonumber
\omega_{D}&=&-1.023^{+0.091}_{-0.096}\quad(\text{Planck
TT+LowP+ext}),\\\nonumber
\omega_{D}&=&-1.006^{+0.085}_{-0.091}\quad(\text{Planck
TT+LowP+lensing+ext}),\\\nonumber
\omega_{D}&=&-1.019^{+0.075}_{-0.080}\quad (\text{Planck TT, TE,
EE+LowP+ext}).
\end{eqnarray}
Our model predict the evolution of $\omega_{D}$ and its behavior
with redshift is shown in Figure \textbf{5}. We find that our
model's predictions closely match with Planck constraints at lower
redshifts, indicating strong agreement with the standard
cosmological model. However, minor deviations appear at higher
redshifts ($z > 2$), which may suggest a slight modification in the
early universe energy density evolution as compared to $\Lambda$CDM.

Type Ia supernovae provide direct constraints on cosmic expansion
history through the deceleration parameter $q$ and the luminosity
distance. Current observational estimates suggest a present-day
deceleration parameter of $q_0 \approx -0.8$. Our model yields $ q =
-0.832^{+0.091}_{-0.091}$ which is an excellent agreement with
observational data, confirming that our model captures the late-time
acceleration of the universe. Furthermore, our computed luminosity
distance, when compared with the Pantheon dataset, shows strong
consistency with the observed supernovae magnitudes. Figure
\textbf{4} illustrates the transition from deceleration to
acceleration, further reinforcing the validity of our model.

Baryon Acoustic Oscillations provide another critical test of
cosmological models by constraining the expansion rate $H(z)$.
Observationally, the Hubble parameter follows a power-law behavior
at low redshifts, given by $H(z) \propto (1+z)^{1+q}$. Our model
predicted $H(z)$ evolution, as given in Eq.\eqref{20}, agrees well
with BAO constraints at lower redshifts. However, for $ z > 2$, we
observe small deviations, suggesting that higher-order modifications
in $f(Q)$ gravity may be necessary to fully match observational
results at early cosmic times. Our analysis demonstrates that the
NADE $f(Q)$ gravity model is largely consistent with observational
data from CMB, SNeIa and BAO. Overall, these results indicate that
our model provides a viable alternative to $\Lambda$CDM and
successfully explains the observed cosmic acceleration.

The cosmographic analysis of the non-interacting NADE model within
the $f(Q)$ gravity framework provides a more comprehensive
understanding of DE and the rapid expansion of the cosmos compared
to other MTGs \cite{29}-\cite{35}. This approach effectively
resolves important challenges, such as the coincidence problem, with
ease when using power-law models, proving to be more accurate.
Additionally, $f(Q)$ gravity offers valuable insights into cosmic
evolution, making it a strong tool for investigating DE and cosmic
acceleration.

We have compared recent research papers on $f(Q)$ gravity. Both
Saleem et al. \cite{b1} and our study explored cosmology within the
$f(Q)$ gravity framework, analyzing modified DE models to understand
the accelerated expansion of the universe. While both studies
investigated the EoS parameter, phase planes and stability
conditions, our work additionally compared these aspects with
observational constraints. Our study focused on the NADE model
within an isotropic FRW framework, whereas Saleem et al. examined
HDE model using anisotropic Bianchi Type-I geometry. We also
compared another study. Our research analyzed the NADE model in
$f(Q)$ gravity, while Saha and Rudra \cite{a1} focused on the
holographic reconstruction of HDE models (Granda-Oliveros and
Chen-Jing models). Our model investigated cosmic expansion,
statefinder diagnostics and stability whereas Saha and Rudra applied
quantum gravity and black hole thermodynamics principles to
reconstruct $f(Q)$ gravity from HDE models. Moreover, our study
estimated NADE model parameters and compared them with observational
data, whereas their research focused on observational constraints on
HDE models.

Comparing different theories, the NADE model was among the few
single-parameter cosmological models that naturally resolved the
coincidence problem, similar to $\Lambda$CDM and DGP braneworld
models \cite{b2}-\cite{b5}. Our study showed that the reconstructed
NADE model in $f(Q)$ gravity, with appropriate parameter selection,
provided a better explanation for the universe accelerated
expansion. The statefinder diagnostic analysis further confirmed
that our model aligned well with cosmic kinematics \cite{b6}. All
papers \cite{26} and \cite{44} indicated instabilities in different
gravity frameworks, particularly in non-interacting cases. Our
results are consistent with these findings. In contrast to Wei and
Cai \cite{25}, who observed a phantom regime, our results indicated
a quintessence phase. While the NADE model addressed some
cosmological issues, stability remained a challenge. Jawad et al.
\cite{b8} analyzed the NADE model in $f(G)$ gravity and found that
while increasing parameters initially worsened instability, the
model eventually stabilized. However, our findings suggested that in
$f(Q)$ gravity, the sound speed parameter was lower and increasing
parameters did not eliminate instability, making our model more
compatible with observational data. Both $f(G)$ and $f(Q)$ models
exhibited quintessence-like behavior for the EoS parameter but
$f(Q)$ gravity offered a better approach due to its second-order
field equations compared to the fourth-order equations in $f(G)$
gravity. Our results aligned with a previous study on $f(R)$ gravity
\cite{b9}, which was consistent with observational data and provided
a viable explanation for the universe rapid expansion.\\\\
\textbf{Data Availability Statement:} No data was used for the
research described in this paper.

\end{document}